\documentstyle[epsf2]{laa}

\newcommand{\grp}   {${\rlap.}^{\circ}$}
\newcommand{\pri}   {${\rlap.}^{\prime \prime}$}
\newcommand{\rl}    {${\rlap.}^{s}$}

\newcommand{\grs}    {GRS 1734$-$292}
\newcommand{\nvss}   {NVSS J173728$-$290802}

\newcommand{\ltsima} {$\; \buildrel < \over \sim \;$}
\newcommand{\simlt}  {\lower.5ex\hbox{\ltsima}}            
\newcommand{\gtsima} {$\; \buildrel > \over \sim \;$}
\newcommand{\simgt}  {\lower.5ex\hbox{\gtsima}}            

\begin{document}

\thesaurus{03(13.25.2;13.18.1;11.19.1;11.09.1)}

\title{The hard X-ray source \grs: \\ 
a Seyfert 1 galaxy behind the Galactic Center}

\author{J. Mart\'{\i}\inst{1}
\and I.F. Mirabel\inst{1}
\and S. Chaty\inst{1}
\and L.F. Rodr\'{\i}guez\inst{2}
}
\institute{
CEA/DSM/DAPNIA/Service d'Astrophysique, Centre d'\'Etudes de Saclay,
91191 Gif/Yvette, France
\and
Instituto de Astronom\'{\i}a, UNAM, Apdo. Postal 70-264,
04510 M\'exico D.F., Mexico}

\offprints{J. Mart\'{\i}}

\date{Received 22 August 1997; Accepted 18 September 1997}

\maketitle

\begin{abstract}

The radio, infrared and optical counterparts of the hard X-ray source 
\grs\ ($l^{II}$=358\grp 9, $b^{II}$=+1\grp 4) are reported. 
The optical spectrum exhibits broad ($>1000$ km s$^{-1}$) H$\alpha$ 
emission at a redshift of 0.0214$\pm$0.0005. 
The radio counterpart is a double-sided synchrotron jet of 5 arcsec, 
which at a distance of 87 Mpc corresponds to a size of 2 kpc. 
The multiwavelength observations of \grs\ indicate that this X-ray source 
is a Seyfert 1 galaxy behind $6\pm1$ magnitudes of visual absorption.

\end{abstract}


\thesaurus{03(13.25.2;13.18.1;11.19.1;11.09.1)}

\section{Introduction}

The observations reported here are part of our ongoing 
project for the identification of radio, optical and infrared counterparts of
hard X-ray sources detected by the
GRANAT satellite in the Galactic Center region (Goldwurm et al. 1994). The
original motivation of this search is based on the fact that
previous identifications of GRANAT sources
have yielded to the discovery of the so called galactic microquasars, i.e., systems
whose physics is regarded as a scaled-down version
of the same processes (outbursts, jets, disks, etc.) occurring in
extragalactic quasars and active galaxies (Falcke \& Biermann 1996).
Their best representative examples known so far include
1E 1740.7$-$2942 (Mirabel et al. 1992), 
GRS 1758$-$258 (Rodr\'{\i}guez et al. 1992) 
and GRS 1915+105 (Mirabel \& Rodr\'{\i}guez 1994). A complete
account of our recent radio observations will be reported in a future
paper (Mart\'{\i} et al. 1998), while here we intend 
to discuss the particular case of \grs.  

The original target source \grs\ was first detected by Sunyaev (1990)
as a previously unknown hard X-ray emitter in the Galactic Center direction.  
The coded mask imaging spectrometer ART-P on board of GRANAT was used in this
discovery. The spectrum between 4-20 keV was well described by a hard
power-law, with a photon index of about $-2$  and  
a total hydrogen column density of $\sim6\times10^{22}$ cm$^{-2}$ 
(Pavlinsky et al. 1994). From the same authors, the corresponding luminosity for a 8.5 kpc distance
was estimated to be $8\times 10^{36}$ erg s$^{-1}$, and the
best ART-P position was known with a 90\% confidence radius of $95^{\prime\prime}$. 
Two years later, on 1992, \grs\ experienced   
a hard X-ray outburst from $<20$ mCrab to 36 mCrab in the 40-400 keV band. This was 
detected by the SIGMA telescope,
also on board of GRANAT, and  
both the rise and the later decline took place in a matter of a few days 
(Churazov et al. 1992). No additional outbursts have been
reported since then. 

From observations in 1995 by
Barret \& Grindlay (1996), a ROSAT-HRI source was proposed 
as the soft X-ray counterpart of \grs\ with a positional accuracy
of a few arcsec. Unexpectedly, we found that a much better position
for the \grs\ candidate could be obtained from the public databases of the 
NRAO VLA sky survey (NVSS), at the wavelength of 20 cm
(Condon et al. 1993).
Indeed, the inspection of the NVSS field of \grs, soon after its release, clearly
revealed to our surprise the presence of a strong compact radio source
well within both the ART-P and the ROSAT error boxes. 
The corresponding radio source designation is \nvss, with its flux density
being at the 48 mJy level. The a priori probability of
finding such a strong radio source within the
ROSAT error box is as small as $\sim2\times10^{-5}$.
This clearly implies that the X-ray and the radio source are
almost certainly related or identical, and the same is also very likely 
to be true for the GRANAT source. 
The position of the NVSS object was available 
with sub-arcsec accuracy and an exhaustive multi-wavelength campaign (optical, infrared and
radio) was soon started based on the accurate radio position. 
The present paper will deal with the observational
evidence, collected during this campaign, that suggests a Seyfert 1 interpretation
for both the NVSS, ROSAT and GRANAT sources. 

\section{Radio continuum observations and results}

\begin{figure*}
\plotfiddle{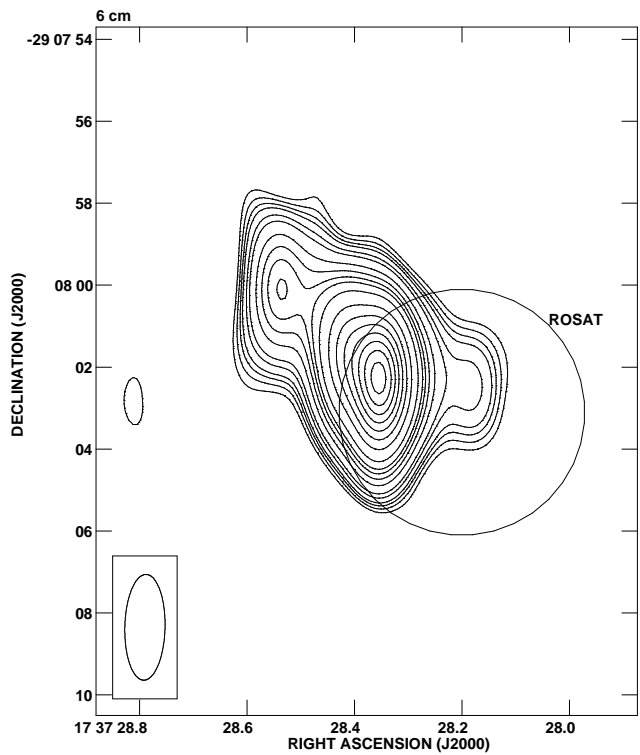}{4.0cm}{0}{45}{45}{-280}{-180}
\plotfiddle{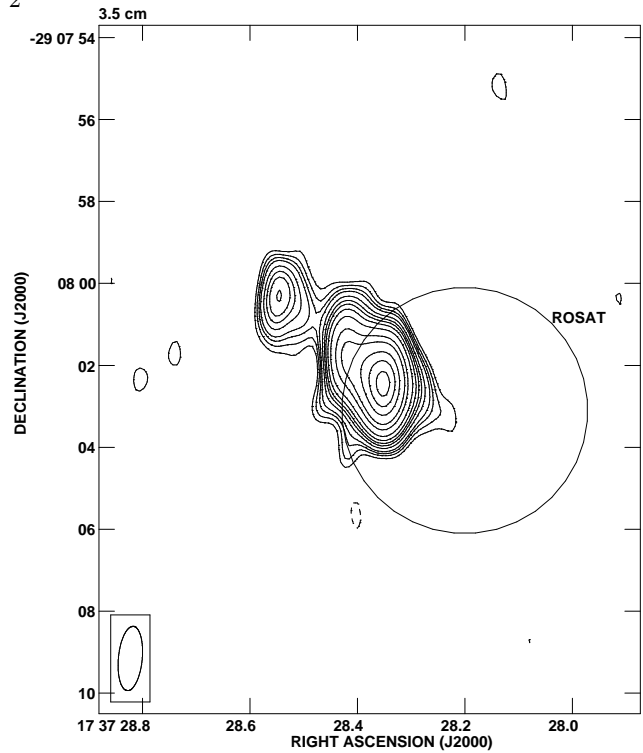}{4.0cm}{0}{45}{45}{-20}{-53}
\caption{\label{maps} Uniform weight maps of \nvss\ at both 6 cm (left)
and 3.5 cm (right) showing the jet-like
structure of this radio source.
The ROSAT 90\% confidence error circle is also indicated in both images.
Contours at 6 cm are $-3$, 3, 4, 5, 6, 8, 10, 15, 20, 25, 30, 40, 60, 80, 100, 120,
140, 160 and 180 times 0.063 mJy beam$^{-1}$, the rms noise.
Contours at 3.5 cm are $-3$, 3, 4, 5, 6, 8, 10, 12, 15, 20, 30, 40, 60, 80 and 100 times
0.053 mJy beam$^{-1}$, the rms noise.
The corresponding synthesized beams are
2\pri 58$\times$0\pri 98 with position angle $-$1\grp4 at 6 cm, and
1\pri 57$\times$0\pri 58 with position angle $-$5\grp5 at 3.5 cm, respectively.
}
\end{figure*}


In a first step, the promising candidate \nvss\ was extensively observed
with the Very Large Array (VLA) interferometer of 
NRAO\footnote{The NRAO is operated by
Associated Universities, Inc., under cooperative agreement with the
USA National Science Foundation.}
in a search for short term radio variability that could confirm its microquasar nature.
The array was most of the time in B configuration. 
The data were processed following standard procedures within the
AIPS software package of NRAO, with 3C 286 and 1751$-$253 being the amplitude 
and phase calibrator, respectively.
The results of our radio monitoring of \nvss\ are summarized in Table \ref{moni},
where the flux density at several wavelengths is listed
for the different dates of observation. 
Some older radio measurements of \nvss\ also quoted in 
Table \ref{moni} could be retrieved from
the literature and the VLA archive database.

\begin{table}
\caption[]{Radio observations of \nvss} \label{moni}
\begin{tabular}{cccc}
\hline
Date        & Julian Day      & $\lambda$    &  Flux Density     \\
            & (JD$-$2400000)  &   (cm)       &  (mJy)            \\
\hline
05 May 1980$^{(1)}$ & 44364.8         &    6         &  $18.8\pm0.3$       \\
       1989$^{(2)}$ &   $-$           &   20         &   56                \\
23 Oct 1993$^{(3)}$ & 49184           &   20         &  $48\pm2$           \\
28 Mar 1997~~~      & 50536.0         &   20         &  $63\pm1$           \\
                    &                 &    6         &  $22.8\pm0.2$       \\
10 Apr 1997~~~      & 50549.0         &   21         &  $57\pm1$           \\
                    &                 &    6         &  $23.7\pm0.2$       \\
                    &                 &   3.5        &  $15.1\pm0.2$       \\
                    &                 &   2.0        &  $8.7\pm0.6$        \\
19 Apr 1997~~~      & 50557.9         &    6         &  $23.9\pm0.2$       \\
                    &                 &   3.5        &  $15.7\pm0.2$       \\
04 May 1997~~~      & 50572.9         &    6         &  $24.0\pm0.3$       \\
                    &                 &   3.5        &  $14.7\pm0.4$       \\
\hline
\end{tabular}
~\\
(1) VLA Archive Database; observer: Sramek R. \\
(2) Helfand et al. 1992, ApJSS, 80, 211 \\
(3) NVSS maps; Condon et al. 1996 (in preparation) \\
\end{table}

In all VLA observations (specially at 6, 3.5 and 2.0 cm) the source appeared resolved
with a clear bipolar jet-like structure. From the 3.5 cm map, the J2000 
position of the central core is found to be  
$\alpha=17^{h}37^{m}$28\rl 35
and 
$\delta=-29^{\circ}08^{\prime}$02\pri 5 ($l^{II}=$358\grp 9, $b^{II}=$+1\grp 4),
with an uncertainty of about 0\pri 1 in each coordinate. 
Contrary to our first expectations,
no significant proper motion in the jet condensations, nor day to day
variability in the source flux density, became detectable in a reliable way. 
In view of that, we decided
to concatenate the $(u,v)$ data of the highest quality sessions
in order to obtain good maps of the \nvss\ radio jets.
The resulting images are presented in Fig. \ref{maps}. The strongest 
jet-like feature emanates in the NE direction and is $\sim3^{\prime\prime}$ extended,
while a weaker $\sim2^{\prime\prime}$ counterjet is also evident.  

Since our VLA monitoring lasted for about five weeks
and there was no positional change in the jet condensations 
larger than $\sim$0\pri 2, the corresponding proper motion upper limit
is about 5 mas d$^{-1}$.
We soon considered all these facts as a first indication that the
object we were studying did not behave as expected from a
microquasar source. 

\begin{figure}
\plotfiddle{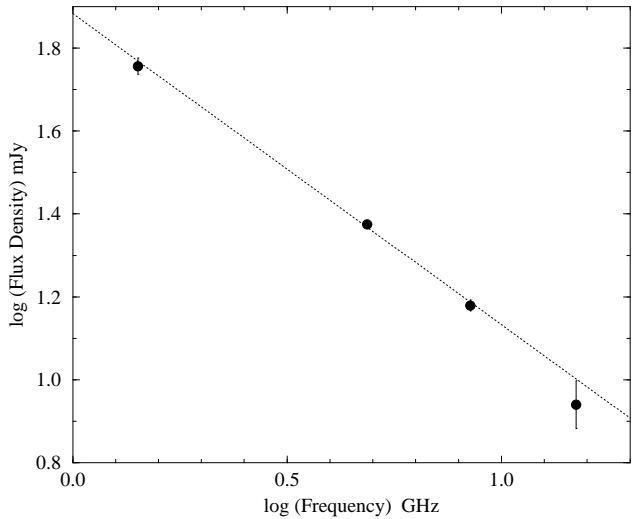}{6.0cm}{0}{50}{50}{-145}{-30}
\caption{\label{lcxu} The non-thermal radio spectrum of \nvss\ from decimetric to centimetric
wavelengths as observed with the VLA on 1997 April 10. The dotted line 
corresponds to a power law fit with the parameters described in the text.}
\end{figure}

On 1997 April 10, the frequency coverage of the observations was wide enough
to measure the spectral properties of the source radio emission. A typical
non-thermal radio spectrum was observed and we present it in Fig. \ref{lcxu}.
A power law fit indicates that the spectrum is well described by
$S_{\nu}=(76\pm4$~mJy~$)(\nu/$GHz$)^{-0.75\pm0.03}$.

\section{Infrared and optical observations and results}

The radio position of \nvss\ was observed at both  
infrared and optical wavelengths using different ESO 
telescopes\footnote{Based
on observations collected at the European Southern Observatory,
La Silla, Chile.}. 
All frames were reduced using standard procedures 
based on the IRAF image processing system.

\subsection{Imaging}

\begin{figure}
\plotfiddle{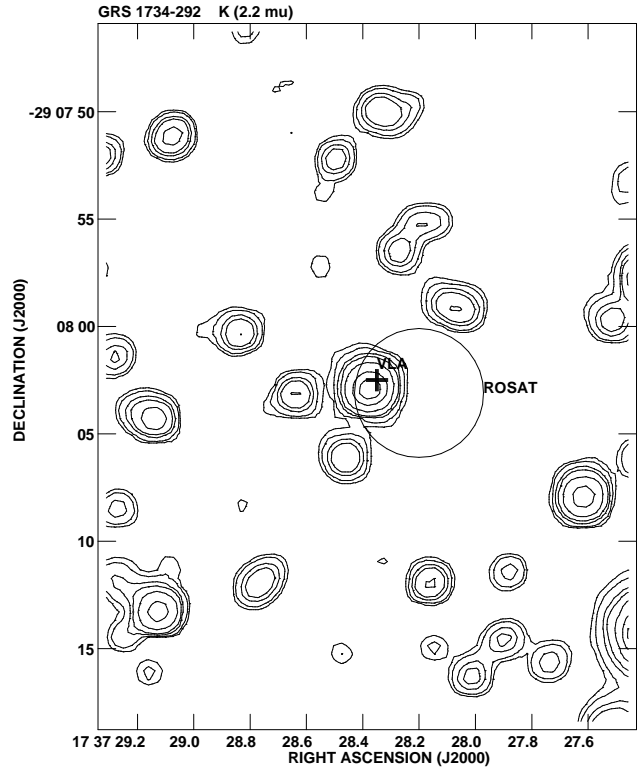}{9.0cm}{0}{45}{45}{-140}{-50}
\caption{\label{finderk} Finding chart of the \nvss\
infrared counterpart in the K band. 
The little cross represents the accurate VLA radio position
and the circle is the ROSAT error box at the 90\% confidence level.
The proposed counterpart is the only object consistent with
both the radio and X-ray positions.}
\end{figure}

\begin{figure}
\plotfiddle{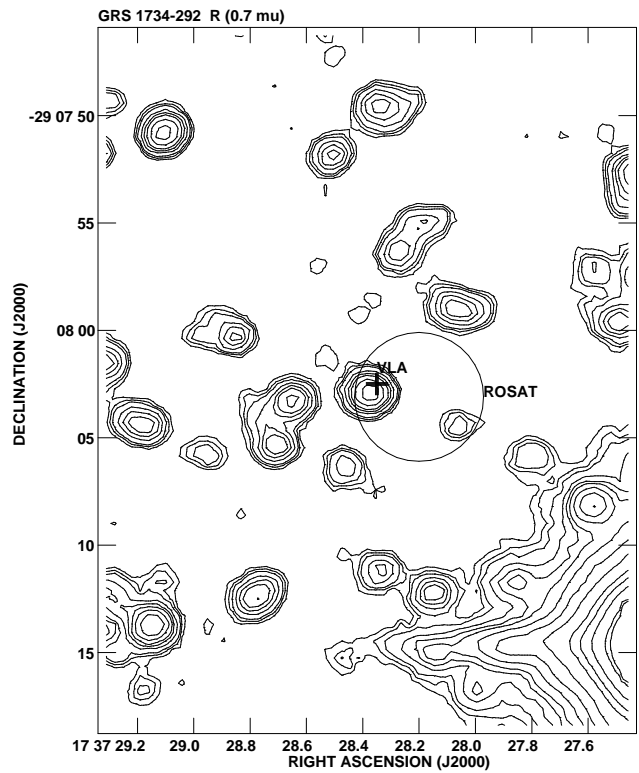}{9.0cm}{0}{45}{45}{-140}{-50}
\caption{\label{finderr} Finding chart of the \nvss\
optical counterpart in the R band with the ROSAT and VLA positions
also indicated. The field is exactly the same as in Fig. \ref{finderk}}
\end{figure}

Imaging infrared observations in the J and K bands were carried
out with the IRAC2b camera mounted at the
F/35 photometer adapter of the 2.2 m telescope. 
We observed on four nights from 1997 March 23 to
April 3. The infrared counterpart of \nvss\ was preliminarily
identified by measuring offsets from
a nearby bright star.
Wide field CCD images in the V, R and I bands were later
obtained with the Danish 1.54 m telescope on 1997 April 10,
using the DFOSC camera whose scale is 0\pri 40 pixel$^{-1}$. An
accurate astrometrical analysis of these optical images
was carried out using nine reference stars from the
Guide Star Catalogue (Taff et al. 1990), thus
confirming our previous infrared identification. The total
offset between the radio and optical position was found to be
about 0\pri 6. This is well within the astrometrical errors of the
fit (rms$\sim$0\pri 4) and also well inside the ROSAT error circle.
We therefore conclude that our identification is correct
and that the optical/infrared counterpart found is the same
object as \grs, \nvss\ and the ROSAT source.
Finding charts at K and R bands to assist in future
observations are shown in Figs. \ref{finderk} and \ref{finderr}.
The two infrared sources IRAS 17342$-$2908
and 358.83+1.39 proposed by Cherepashchuk et al. (1994)
as counterpart candidates are not consistent with our identification.

On the other hand, our identified source did not display
clear evidences of photometric variability  
during both the infrared and optical imaging observations. Although we only
had one night at the 1.54 m telescope, the lack of variability in the
optical can be established from the similar $R$ band magnitude derived
during the spectroscopic session described below. 
The photometry of the source is summarized in Table \ref{photo}.

\begin{table}
\caption[]{Magnitudes of the \grs\ counterpart} \label{photo}
\begin{tabular}{cccc}
\hline
Filter      & Observation Date  & Telescope      & Magnitude     \\
\hline
  V         & 1997 April 10     & 1.54 m + DFOSC &  $21.0\pm0.3$ \\
  R         & 1997 April 10     & 1.54 m + DFOSC &  $18.3\pm0.1$ \\
  I         & 1997 April 10     & 1.54 m + DFOSC &  $16.8\pm0.1$ \\
  J         & Average all dates & 2.2 m + IRAC2b &  $13.7\pm0.1$ \\
  K         & Average all dates & 2.2 m + IRAC2b &  $11.1\pm0.1$ \\ 
\hline
\end{tabular}
\end{table}

\subsection{Spectroscopy}

\begin{figure}
\plotfiddle{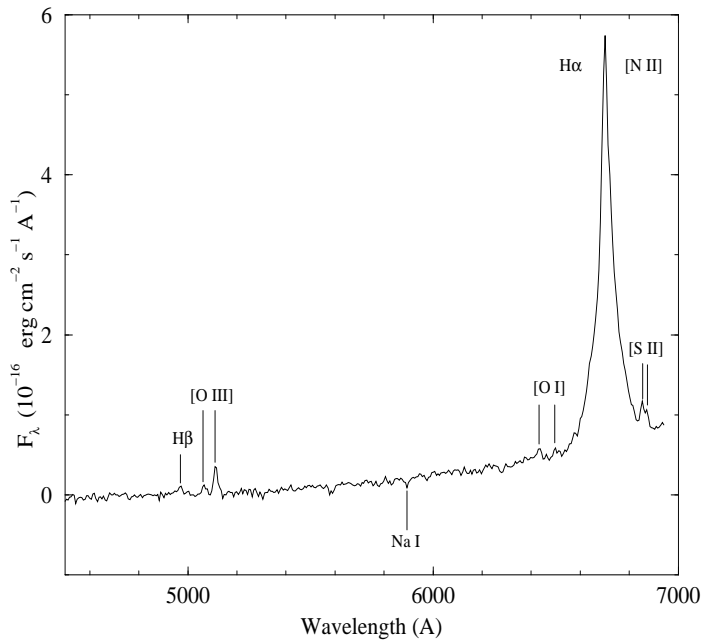}{7.5cm}{0}{55}{60}{-165}{-35}
\caption{\label{halpha} Optical broad band spectrum of \grs\
obtained on 1997 March 12 with the EFOSC1 instrument
at the 3.6 m ESO telescope.}
\end{figure}

Broad band spectroscopic observations of the \grs\ optical counterpart
were also carried out with EFOSC1 on the 3.6 m ESO telescope, using 
the B300 grism whose dispersion is 2.0 \AA\ pixel$^{-1}$.
As shown in Fig. \ref{halpha}, they revealed 
that the optical spectrum is completely dominated by
strong and very broad emission from the blended H$\alpha$ and [NII] lines.
Other emission lines from [OI], [OIII]
and [SII] are also identifiable. A consistent redshift measurement is obtained
from all of them, with our best estimate being $z=0.0214\pm0.0005$. 

The full width zero intensity of the blended H$\alpha$ and [NII] is extremely broad, 
nearly 450 \AA\ or $\sim20000$ km s$^{-1}$. 
Their total flux in emission is $3.8\times 10^{-14}$ erg s$^{-1}$ cm$^{-2}$. 
We have attempted to deblend
these three lines using Gaussian components at the expected
wavelengths for $z=0.0214$. The results obtained are given in Table \ref{abso}.
This table also includes information 
on the other spectral lines most evident in the \grs\ spectrum.
We wish to point out that the deblending
procedure may not be completely reliable here because the
H$\alpha$ and [N II] lines are very difficult to separate. 
In particular, the full width half maximum (FWHM) derived for the forbidden [N II] components (often
narrower than $\sim$1000 km s$^{-1}$ in Seyferts) 
seems to be unusually high, and perhaps we are underestimating the H$\alpha$
emission. It is also possible that H$\alpha$ has an even broader component
that our deblending fit is not accounting for. 

\begin{table*}
\caption[]{Main lines in the \grs\ optical spectrum} \label{abso}
\begin{tabular}{lcrrrrl}
\hline
Line        & Observed wavelength & Observed flux       & FWHM &  FWHM         &  EW  & Notes     \\
            &   (\AA)             & (erg s$^{-1}$ cm$^{-2}$) & (\AA)& (km s$^{-1}$) & (\AA) &    \\
\hline
$[$S II$]$  & 6853.3 &  $4.8\times 10^{-16}$  & 15   &   660  & $-6$   &   \\
            & 6871.7 &  $2.5\times 10^{-16}$  & 12   &   530  & $-3$   & \\
H$\alpha$   & 6701.5 &  $8.3\times 10^{-15}$  & 30   &  1340  & $-120$ & Peak wavelength   \\
$[$N II$]$  & 6688.1 &  $1.5\times 10^{-14}$  & 123  &  5510  & $-216$ & Assumed wavelength for deblending \\
            & 6723.9 &  $1.5\times 10^{-14}$  & 104  &  4640  & $-200$ & Assumed wavelength for deblending \\ 
$[$O I$]$   & 6432.4 &  $4.2\times 10^{-16}$  & 32   &  1490  & $-10$  &   \\
            & 6503.5 &  $4.2\times 10^{-16}$  & 35   &  1610  & $-9$   &   \\
Na I D      & 5893.5 &  $-1.8\times 10^{-16}$ & 17   &   870  &  $9$   & Interstellar line    \\
$[$O III$]$ & 5113.5 &  $6.8\times 10^{-16}$  & 17   &  1000  & $-400$ & Continuum very weak  \\
            & 5066.5 &  $2.1\times 10^{-16}$  & 18   &  1070  & $-172$ & Continuum very weak  \\
H$\beta$    & 4966.0 &  $2.8\times 10^{-16}$  & 31   &  1870  & $-555$ & Continuum very weak  \\
\hline
\end{tabular}
\end{table*}

\section{The column density towards GRS 1734$-$292}

In this section, we undertake a comparison study of 
the total absorption column density $N($H$)$
towards \grs\ derived by three independent methods. A mean value is finally
adopted in order to deredden the photometric magnitudes of the previous section.
This will help us later to 
estimate the unabsorbed optical luminosity of \grs\ when discussing
its physical nature.

\subsection{An upper limit to $N($H$)$ from the sodium interstellar absorption line}

The spectrum in Fig. \ref{halpha} displays an absorption feature at 
5893.5 \AA\ (see Table \ref{abso}). This can be interpreted as an unresolved detection
of the two Na D interstellar absorption lines. The Na D lines are expected to be at
5890 and 5896 \AA, respectively. The absorption feature mentioned is well located
at the middle point between these two
wavelengths. The identification is thus convincing, although our resolution is 
not good enough to distinguish the two components separately. 

The intensity of the Na D absorption feature can be used to estimate both 
the extinction and the distance for objects
within the Galaxy. The corresponding relationship is given by:
$$ A_V = 3.8  EW,$$ 
where $EW$ is the mean equivalent width in \AA\ of the two Na D-lines,
and a 1.9 mag kpc$^{-1}$ absorption of optical light near the Galactic Plane
has been assumed (Allen 1973). 
For extragalactic sources, the Na lines only provide an indication of the
length of the line of sight within our own Galaxy. This is, of course,
making the reasonable assumption  
that all the Na absorption is produced inside the Milky Way.

In our case, we are not able to resolve the Na lines and only an upper limit to the
mean equivalent width is available from observation ($EW \leq 9$ \AA). This implies
from the previous equation that
the extinction towards \grs\ should be
$A_V \leq 34.2$ mag, and consequently 
the line of sight towards \grs\ intercepts less than 18 kpc of 
gas and dust within our Galaxy. 
From the relationship by Predehl \& Schmitt (1995):
$$A_V = 0.56 \left[\frac{N(H)}{10^{21}~cm^{-2}}\right] +0.23,$$
this corresponds to a total hydrogen column density of about
$N($H$) \leq 6.1\times 10^{22}$ cm$^{-2}$. 

This upper limit is consistent in order of magnitude with the
rough estimate $\sim6\times10^{22}$ cm$^{-2}$, or $A_V\sim 34$ mag,
derived by Pavlinsky et al. (1994) from X-ray model fitting with their ART-P data.
However, the fact that an optical counterpart has
been found for \grs\ is difficult to reconcile with an optical extinction
of more than thirty magnitudes. Therefore, the ART-P column density
is likely to be strongly overestimated and we will show below
that this is certainly the case.  

\subsection{$N($H$)$ estimated from neutral hydrogen absorption}

A H I absorption experiment at 21 cm  was carried out with the VLA
on 1997 April 10 and it was also reduced using standard
AIPS procedures.
The resulting spectrum (not shown here)
was not of high signal to noise ratio. The only absorption feature
detected was localized at $3.6\pm0.1$ km s$^{-1}$ LSR velocity,
with an estimated peak opacity value of $\tau_0=1.10\pm0.02$. This implies
that most of the absorption is produced in the Saggittarius arm.
The FWHM was $21.1\pm0.3$ km s$^{-1}$. The corresponding
column density of H I along the line of sight to \nvss\ can be
expressed as $N($H I$)=(4.2\pm0.2)\times10^{21} (T_s/100$~K$)$ cm$^{-2}$, where
$T_s$ is the hydrogen spin temperature. Using the canonical value $T_s=125$ K,
we estimate that $N($H I$)=(5.3\pm0.3)\times10^{21}$ cm$^{-2}$.

Since \grs\ is close to the Galactic Center direction,
$N($H$)$ should include another important contribution from
the metals associated to the abundant molecular hydrogen component $N($H$_2)$
in addition to the atomic species.
In order to derive $N($H$_2)$, we have used the
Columbia $^{12}$CO (J=1-0) survey by Dame et al. (1987)
together with the empirical relation of $N($H$_2)$ with the
integrated CO line intensity. This relation can be expressed
as $N($H$_2)=3.6\times 10^{20}\int T($CO$)dv$ (Scoville et al. 1987).
By interpolating the Columbia survey
at the \grs\ position, we find one single
emission component at a LSR velocity of $-5\pm1$ km s$^{-1}$.
A Gaussian fit to this line yields a peak temperature of $0.29\pm0.02$ K
with a FWHM of $22\pm3$ km s$^{-1}$.
The corresponding value of $N($H$_2)$ is thus $(2.4\pm0.4) \times 10^{21}$ cm$^{-2}$.
 
By combining the H I and CO information,
the total absorbing column density in the \grs\ direction
can now be found from $N($H$)=N($H I$)+2 N($H$_2)$. The final result is
$N($H$)= 1.0\pm0.1 \times 10^{22}$ cm$^{-2}$, equivalent to a visual
extinction of $A_V=5.8\pm0.5$ magnitudes 
using again the Predehl \& Schmitt (1995) relation.
It is important to mention here that recent studies by Dahmen et al. (1996)
suggest that the conversion factor between the CO emission and the H$_2$ column
density may be overestimated by an order of magnitude. If this is the case,
the $A_V$ value derived above should be consequently revised. Nevertheless, a reliable lower
limit of $A_V>3.2$ magnitudes can be established from the H I contribution alone.  

\subsection{$N($H$)$ estimated from the Balmer decrement}
 
The extinction and the hydrogen column density towards \grs\ can also be 
independently estimated from the measured line ratio H$\alpha$/H$\beta$
in the optical spectrum and using Table \ref{abso} values.
Following Miller \& Mathews (1972), the H$\alpha$/H$\beta$
relationship with galactic extinction can be expressed as:
$$A_B=8.5 \log{((H\alpha/H\beta)/3.0)},$$
while absorption at other bands can be easily computed using
the following parameterized reddening curve:
$$A_{\lambda}=0.74 \lambda^{-1} -0.34,$$
where $\lambda$ is the central band wavelength expressed in microns.

The H$\alpha$ flux in \grs\ is, of course, less than the total blended
emission of the H$\alpha$ and [N II] lines (H$\alpha < 3.8\times 10^{-14}$ erg s$^{-1}$ cm$^{-2}$).
This implies that H$\alpha$/H$\beta < 136$, and we confidently estimate that
$A_V<10.5$ mag. Furthermore, if 
the deblending procedure in Table \ref{abso} was appropriate,
we would find more precisely that H$\alpha$/H$\beta\simeq30$. Such a line ratio then yields
$A_V=6.3$ mag, with a formal likely uncertainty of $\pm0.5$ mag. This
final absorption estimate translates into $N($H$)=(1.1\pm0.1)\times 10^{22}$ cm$^{-2}$.

\vspace{0.5cm}
Summarizing this section, all three independent used methods seem to provide consistent results.
This agreement may be further tested in the future by carrying out 
additional spectroscopic optical and radio observations with higher resolution
and sensitivity. In the following, we will adopt $A_V=6\pm1$ mag, or equivalently 
$N($H$)=(1.0\pm0.2)\times10^{22}$ cm$^{-2}$, as a compromise mean value for discussion purposes.
Such an amount of visual extinction is indeed 
a reasonable result at 1\grp 4 of galactic latitude. The mean values of $A_V$ 
as a function of $b^{II}$, and close to the
Galactic Center direction, have been studied 
for instance by Catchpole et al. (1990). The statistical analysis 
of these authors using colour-magnitude diagrams does provide $A_V\sim 5$ mag for galactic latitudes
in the 1\grp0-1\grp5 range, i.e., where \grs\ is located.

\section{Discussion}

\begin{table*}
\caption[]{The main physical parameters of \grs} \label{param}
\begin{tabular}{lll}
\hline
Parameter          &   Value                                          &   Notes                      \\
\hline
Redshift           & $z=0.0214\pm0.0005$                              &                              \\
Distance           & $D=87$ Mpc                                       & $H_0=75$ km s$^{-1}$         \\
Jet size           & $l_{jet}\simeq2.1$ kpc                           &                              \\
Visual absorption  & $A_V = 6\pm1$                                    &                              \\
H column density   & $N($H$)=(1.0\pm0.2) \times 10^{22}$ cm$^{-2}$    &                              \\
Radio luminosity   & $L_{rad}\simeq 7\times 10^{39}$ erg s$^{-1}$     & 0.1-100 GHz band             \\
Optical luminosity & $L_{opt}\simeq 2\times 10^{43}$ erg s$^{-1}$     & 4900-9000 \AA\ band          \\
X-ray luminosity   & $L_{X}\simeq 1\times 10^{44}$ erg s$^{-1}$       & 0.5-4.5 keV band             \\
Line luminosity    & $L_{H\alpha} \simeq 5\times 10^{41}$ erg s$^{-1}$     & Deblended value      \\
                   & $L_{H\beta} \simeq 1 \times 10^{41}$ erg s$^{-1}$     &       \\
                   & $L_{[S~II]} \simeq 4 \times 10^{40}$ erg s$^{-1}$     &       \\
                   & $L_{[O~I]} \simeq  6 \times 10^{40}$ erg s$^{-1}$     &       \\
                   & $L_{[O~III]} \simeq 4 \times 10^{41}$ erg s$^{-1}$    &       \\
\hline
\end{tabular}
\end{table*}

The observed redshift of \grs\ corresponds 
to a recession velocity of 6500 km s$^{-1}$. Such a high
value rules out any interpretation based on the systemic velocity of a binary star 
in the Galaxy. For this galactic interpretation, 
one should expect a redshift (or blueshift) of at most \ltsima1000 km s$^{-1}$, i.e.,
the typical kick velocity acquired by the binary system after the supernova
explosion forms the compact companion.  
On the contrary, the simplest way to account for the observed redshift is to
assume that \grs\ lies at a cosmological distance and is, therefore, an extragalactic source. 
Using Hubble's law, the corresponding distance can be estimated as
$D = 65 h^{-1}$ Mpc (where the Hubble constant
is expressed here as $H_0=100h$ km s$^{-1}$ Mpc$^{-1}$ and a Universe
with $\Omega=1$ is assumed).

The spectrum in Fig. \ref{halpha}
is highly reminiscent of a Seyfert 1 Galaxy given the large width 
of permitted lines. For a Seyfert galaxy at a $z=0.0214$ redshift, it should
normally be possible to see some arcsec extended nebulosity if located
at high galactic latitude. The deep R band CCD image
in Fig. \ref{finderr} shows that this is not the case.
The discovered optical counterpart 
appears as an unresolved source and only the galactic nucleus is evident.  
This is possibly due to the optical extinction in the galactic plane ($b^{II}=$1\grp 4).
Although not extremely great ($A_V\simeq6$ mag), it is apparently sufficient to 
prevent any faint nebulosity from being seen in the optical. In the
K band no nebulosity is seen either, and this compactness could mean that  
\grs\ is a nearby quasar instead of a Seyfert 1. However, we believe
that the optical spectrum observed is a very strong
evidence to prefer by now a Seyfert 1 interpretation.

The catalogue of GRANAT sources does not include many examples of
extragalactic objects and our discovery adds a new member to this
scarce group. Only three extragalactic sources have been extensively
detected by the SIGMA telescope on board of GRANAT. They are the quasar
3C 273, the Seyfert 1.5 galaxy NGC 4151, and the radio galaxy Cen A (Bassani
et al. 1993; Jourdain et al. 1993). All of them are characterized by displaying 
clear hard X-ray variability and spectral
evolution in time scales of both years and, in some cases, few days.
In particular, Cen A was observed to decrease its 40-120 keV flux
by a factor of 1.5 within only four days. This behavior compares well with that of
\grs. Our target source is currently accepted as a confirmed variable (Barret \& Grindlay 1996),
and it exhibited a few day time scale variability during its 1992 hard X-ray outburst
(Churazov et al. 1992). On the other hand, the \grs\ spectrum
became extremely hard during this flaring event, and this is remarkably similar
to the spectrum hardening observed by SIGMA in the
NGC 4151 Seyfert during epochs of high photon flux (Bassani et al. 1993). 

There are in addition other observational clues in agreement with a Seyfert 1 galaxy 
scenario. For instance, 
the total radio power derived from
the spectrum in Fig. \ref{lcxu} is $L_{rad} \simeq 4 \times 10^{39} h^{-2}$ erg s$^{-1}$
and, in particular, $P_{21 cm} \simeq 3 \times 10^{29} h^{-2}$ erg s$^{-1}$ Hz$^{-1}$.
This correlates well when plotted in monochromatic 21 cm radio power versus
0.5-4.5 keV X-ray luminosity diagrams for Seyfert 1
galaxies by Wilson (1991). The unabsorbed value of the X-ray output is taken
to be here $L_X \simeq6\times 10^{43} h^{-2}$ erg s$^{-1}$ by extrapolating
the power law fit from Pavlinsky et al. (1994). 

The observed radio jet morphology and spectral index 
are quite common among Seyfert galaxies. The case of \grs\
should be classified as belonging to the L or {\it linear} class
in the Wilson (1991) scheme.
The radio jet size ($l_{jet} \simeq 1.6 h^{-1}$ kpc)
also correlates acceptably well with the radio power from a Seyfert 1 object
(Ulvestad \& Wilson 1984).
The optical luminosity estimated from our broad band spectrum
and VRI photometry using $A_V=6$ mag 
is $L_{opt}\sim 1 \times 10^{43} h^{-2}$ erg s$^{-1}$ in the 4900-9000 \AA\ band.
This is again in good order of magnitude agreement
with expectations based on the radio/optical power correlation
studied by Edelson (1987) for Seyfert galaxies in the CfA sample.
Other correlations that test acceptably well are those involving the
[O III] luminosity versus H$\beta$ luminosity, radio power and
FWHM of [O III] (Lawrence 1987; Whittle 1985). 

We close this discussion by giving in Table \ref{param} the
main physical parameters of \grs, expressed 
for the particular case of $H_0=75$ km s$^{-1}$, and mentioning
that this Seyfert also fits reasonably well the Falcke \& Biermann (1996) scheme for AGNs
when plotted in their diagram of monochromatic radio power versus core disk luminosity. 

\section{Conclusions}

We have presented observations that provide a
very accurate positional identification of the 
radio, infrared and optical counterpart of the GRANAT source \grs. 
The discovered counterpart displays clear evidence of being a   
Seyfert 1 galaxy. The most remarkable properties of the system
are perhaps its clear linear jet-like structure and its  
broad H$\alpha$ emission. 

A redshift measurement yields the value $z=0.0214\pm0.0005$,
thus providing a distance to \grs\ of  
87 Mpc ($H_0=75$ km s$^{-1}$ Mpc$^{-1}$). The column density 
towards \grs\ is also estimated using three different
techniques and an average value of $A_V=6\pm1$ mag is proposed.  
This is equivalent to a hydrogen column density of $N($H$)=(1.0\pm0.2)\times 10^{22}$ cm$^{-2}$.  
The Seyfert 1 nature of \grs\ is additionally confirmed by a satisfactory agreement 
with different well established correlations for Seyfert galaxies. 
We also point out that the hard X-ray behavior of \grs\
is consistent with 
extragalactic sources studied by GRANAT.

\acknowledgements{J.M. acknowledges financial support from a postdoctoral
fellowship of the Spanish Ministerio de Educaci\'on y Ciencia.
LFR acknowledges support from DGAPA, UNAM and CONACyT, Mexico.
We thank C. Lidman who arranged the ESO observations in service mode 
as well as F. Comer\'on who kindly obtained some
of the images. A.S. Wilson,  
J. Paul, J. Lequeux, C. Gouiffes and P.-A. Duc are also acknowledged
for useful comments, help and discussion. Carlos De Breuck is
specially thanked for obtaining the optical spectrum.
This research has made use of the Simbad database,
operated at CDS, Strasbourg, France.
}

\end{document}